\documentclass[10pt,twocolumn,letterpaper]{article}

\usepackage{cvpr}
\usepackage{times}
\usepackage{epsfig}
\usepackage{graphicx}
\usepackage{amsmath}
\usepackage{amssymb}

\usepackage[breaklinks=true,bookmarks=false]{hyperref}

\cvprfinalcopy 

\def\cvprPaperID{****} 

\begin{document}

\title{Audio segmentation based on melodic style with hand-crafted features and with convolutional neural networks}

\author{Amruta Vidwans, Nachiket Deo, Preeti Rao \\ Department of Electrical Engineering, Indian Institute of Technology, Bombay\\ {\tt \{amrutav, ndeo, prao\}@ee.iitb.ac.in}}


\maketitle

\begin{abstract}
We investigate methods for the automatic labeling of the \textit{taan} section, a prominent structural component of the Hindustani \textit{Khayal} vocal concert. The \textit{taan} contains improvised \textit{raga}-based melody rendered in the highly distinctive style of rapid pitch and energy modulations of the voice. We propose computational features that capture these specific high-level characteristics of the singing voice in the polyphonic context. The extracted local features are used to achieve classification at the frame level via a trained multilayer perceptron (MLP) network, followed by grouping and segmentation based on novelty detection. We report high accuracies with reference to musician annotated \textit{taan} sections across artists and concerts. We also compare the performance obtained by the compact specialized features with frame-level classification via a convolutional neural network (CNN) operating directly on audio spectrogram patches for the same task. While the relatively simple architecture we experiment with does not quite attain the classification accuracy of the hand-crafted features, it provides for a performance well above chance with interesting insights about the ability of the network to learn discriminative features effectively from labeled data.
\end{abstract}
\section{Introduction}\label{sec:introduction}
Structural segmentation of concert audio recordings is very useful in music retrieval tasks such as navigation and automatic summarization. It is particularly strongly indicated for Indian classical music where concerts can extend for hours, and commercial audio recordings are rarely annotated, while the performance indeed follow an established structure depending on the genre. \textit{Khayal} vocal music is the single most prominent genre in the Indian classical tradition of Hindustani music. A raga performance in \textit{khayal} has a structure comprised of a number of elements such as the free form introduction (\textit{alap}), the composition (\textit{bandish}), metered improvisation (also, \textit{alap}), rhythmic improvisation (\textit{layakari}) and improvisation involving fast sequences of notes (\textit{taan}) \cite{rao2014overview}. The concert ensemble is made up of the vocalist accompanied by the drone and percussion and sometimes melodic accompaniment such as the harmonium or \textit{sarangi}. As such there are no changes in timbre texture due to the constancy of instrumentation, and harmony is non-existent. The structural elements mentioned earlier occur to various extents in the performance and in different orders depending on the school (\textit{gharana}). Even to the uninitiated (but attentive) listener, the different concert sections appear clearly contrasting in one or the other of the two important dimensions: rhythm and melodic style.  Recently, tempo derived features were used to achieve structural segmentation at the highest time scale on Hindustani instrumental concert audio \cite{PV}.

In this work, our focus is on segmenting sections that are melodically salient i.e. the sequence of melodic phrases or notes is rendered in a characteristic melodic style known as the \textit{taan}. The notes may be articulated in various ways including solfege and the syllables of the lyrics. Most common however is the \textit{akar} \textit{taan}, rendered using only the vowel /a/ (i.e. as melisma). The sequence of notes is relatively fast-paced and regular, produced as skillfully controlled pitch and energy modulations of the singer’s voice similar to vibrato. But unlike the use of vibrato which ornaments a single pitch position in Western music, the cascading notes of the \textit{taan} sketch elaborate melodic contours like ascents and descents over several semitones. The melodic structure is strictly within the \textit{raga} grammar while the step-like regularity in timing brings in a rhythmic element to the improvisation in contrast to the (also improvised) \textit{alap} sections. Apart from showcasing the singer’s musical skills, one or more \textit{taan} sections typically contribute to the climax of a raga performance and therefore serve as prominent markers musicologically.

A broad overview of methods available for structural segmentation is summarized in \cite{Paulus}. Since our task involves the detection and segmentation of a specific named section of the concert, we need to invoke both segmentation and supervised classification methods. Musically motivated features and methods are our chosen approach given their potential for success with limited training data \cite{XSerra}. The challenges to \textit{taan} detection are the polyphonic setting where we want to focus on the vocal signal, and designing distinctive features that are artist and concert independent. Given that pitch modulations are the prime characteristic of \textit{taan}, reliable pitch detection with sufficient time resolution is necessary. Finally, we need to convert the low level analyses to annotation that closely matches with the musician’s labeling of \textit{taan} episodes from a performer’s point of view. Towards these goals, we use a vocal source separation algorithm based on predominant-F0 detection \cite{VRaoMelodyExtract}. Features designed to capture the characteristic of rapid but regular pitch and energy variations of the voice are presented. A frame level classification at 1 s granularity is followed by a grouping stage with the goal of emulating the subjective labeling of \textit{taan} by musicians as extended regions that occur at salient positions in the concert. Finally, we also wish to explore the interesting question of whether the hand-designed features can be replaced by learned features obtained via a CNN applied directly to the polyphonic audio spectra. There has been much recent research interest in automatic feature learning for a variety of audio tasks such as genre and artist classification \cite{AndrewCNN}, chord recognition \cite{ChordRecogBello}, onset detection \cite{ImprovedOnsetCNN}, and structural analysis \cite{Ullrich}.

In the next section, we describe the characteristics of our audio database. This is followed by a discussion of the proposed melodic style features, and the classification and segmentation methods. Finally, we present the experiments and evaluation measures followed by a discussion of the results.

\section{Database Description}\label{Database}
Our audio database consists of 57 khayal vocal concert recordings from commercial CDs partitioned into two distinct sets of 22 single-artist (Pt. Jasraj) concerts, and 35 multi-artist concerts (that do not contain Pt. Jasraj). In both cases a number of different ragas are covered at various tempi. All artists are male. The 22 concert set is treated as the test set with two different training conditions: artist-specific training via leave-one-song-out cross validation, and artist-independent training where the test concert artist is not represented at all in the training data of 35 concerts. In order to achieve realistic training time for the CNN classification in the 22-fold cross-validation, the 22 test concert audios were edited to remove an early sections of each concert audio (where \textit{taan} typically does not occur). Eventually we have 3.5 hours of test audio with the proportion of \textit{taan} in the overall vocal region at 35\%.

\begin{figure}[t]
 \centerline{\framebox{
 \includegraphics[width=\columnwidth,height=4cm]{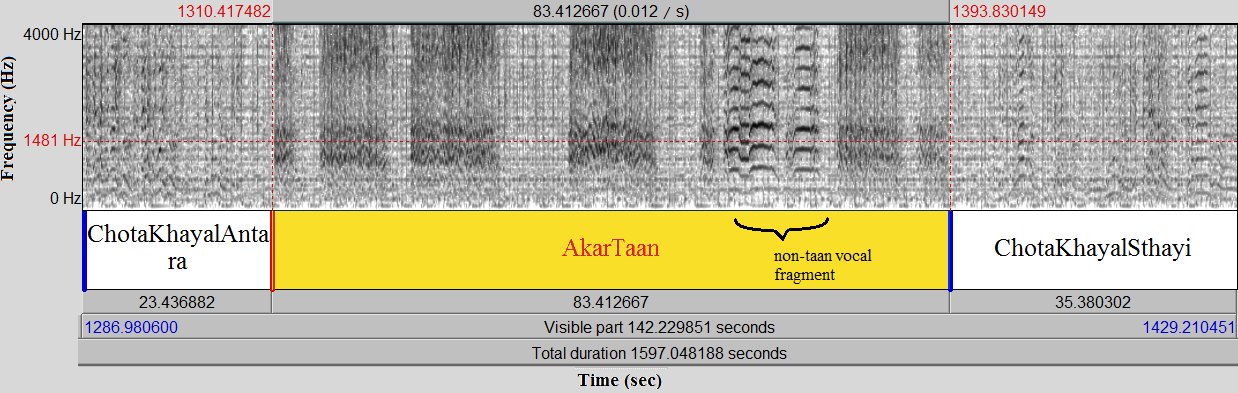}}}
 \caption{Spectrogram of an episode of \textit{akartaan} flanked by other sections in a concert.The labeled \textit{taan} section  shows rapid oscillatory movements of vocal harmonics, interrupted by short non-\textit{taan} movement in-between.}
 \label{fig:PraatSpecImg}
\end{figure}

The labeling of concert sections was carried out by a musician using the PRAAT interface \cite{Praat}. \ref{fig:PraatSpecImg} shows a fragment of labeled audio comprising portions of 3 sections spanning 2.5 minutes of a \textit{khayal} performance. We observe that the single continuous section labeled ‘\textit{akar taan}’ (of duration 85 s) actually comprises of a cluster of \textit{taan} segments separated by instrumental or other regular singing segments. A \textit{taan} segment is easily identified in the audio spectrogram, computed with 40 ms Hamming windowing, by the modulated harmonics in the region of prominent formants (dark region above 800 Hz). Within the labeled \textit{taan} section, the individual \textit{taan} episodes can be as short as 5 s and be separated from each other by up to as much as 20 s. We observed that the musician labeled \textit{taan} based on the perceived intent of the performer i.e. relatively short durations of instrumentals and other vocal styles that occurred sandwiched between \textit{taan} episodes were subsumed by the \textit{taan} label (as in \ref{fig:PraatSpecImg}). For the real-world use case, we would like our automatic system to match the musician’s labeling of the \textit{taan} sections in the concert.

\section{Feature extraction, classification and grouping}\label{sec:FeatExtrClasGrpng}
Given our knowledge about \textit{taan} production and observations of the acoustic signal characteristics, it is clear that the presence of strong pitch modulation is among the distinctive traits of the \textit{taan} style of singing. The required audio pre-processing and feature extraction methods are presented in the following.

\subsection{Vocal attributes extraction}\label{subsec:VocalAttExtr}
The singing voice usually dominates over other instruments in a vocal concert performance in terms of its level and continuity over relatively large temporal extents although the accompaniment of \textit{tabla} and other pitched instruments such as the drone and harmonium are present. The singing voice regions, or vocal spurts, are identified in the audio track using an available singing voice detection system based on timbral and periodicity characteristics of the singing voice as opposed to the instrumentation \cite{VRaoContextAware}. The SVM classifier is trained on a few hours of Hindustani vocal music (different from the database used in the present work). Next, a predominant F0 detector is used to estimate the pitch at 10 ms intervals corresponding to the vocal component \cite{VRaoMelodyExtract}. The pitch detector uses an adaptive analysis window to optimize the time and frequency resolution trade-off in order to track rapid pitch variations. The total harmonic energy in the frequency region below 5 kHz, where the harmonics correspond to the detected pitch, provides an estimate of the vocal energy, also at 10 ms intervals. The purely instrumental regions, as determined by the singing voice detector, are not processed for feature extraction.

\subsection{Pitch and energy modulation features}\label{subsec:Features}
\begin{figure}[t]
 \centerline{\framebox{
 \includegraphics[width=\columnwidth]{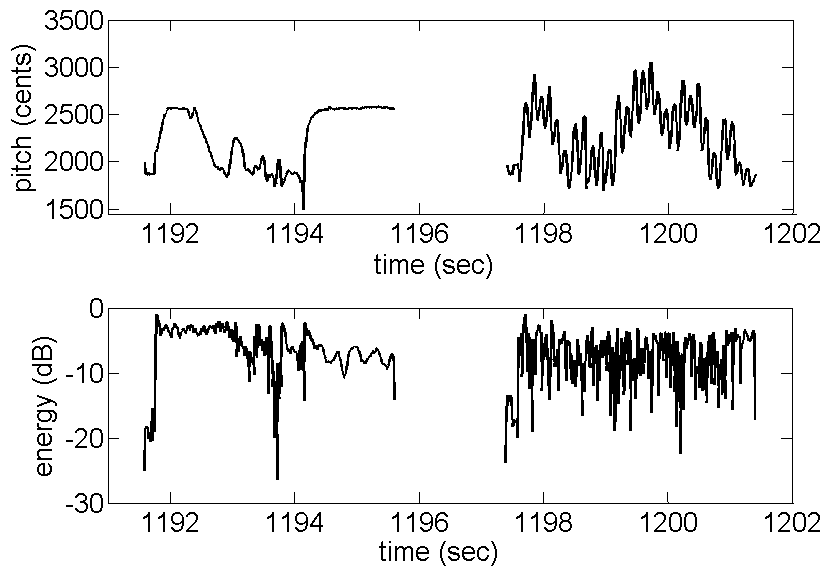}}}
 \caption{Pitch contour in cents (top) and energy in dB (bottom) of a section of concert audio. Non-\textit{taan} movements seen in the first half and taan seen after.}
 \label{fig:PitchEnergyContour}
\end{figure}

The melodic style descriptors are computed in the detected vocal regions only. The pitch values are first converted to a cents scale by normalising with a standard F0 chosen to be 55 Hz. The sampled pitch trajectory within each 1 s analysis frame is mean subtracted where mean refers to the slow trend in melodic shape. The mean smooth trajectory is obtained by a third order polynomial fit to the pitch samples in the frame \cite{ChitraSpringer}. The mean subtracted trajectory is analysed by the 128 point DFT of a sliding window of 1 s duration at 500 ms hop intervals to find the spectrum peak location and height in the region 1-20 Hz. The peak location is an estimate of the pitch modulation rate. It is observed to lie in the 5-10 Hz range irrespective of the underlying tempo of the section in the case of \textit{taan} like movements. The energy computed from the DFT power spectrum in a neighborhood of +/- 1.6 Hz (5 bins) around the peak represents the regularity and strength of the pitch modulation. It was also observed that the overall energy in the voice fluctuated with the pitch modulation. This could be a consequence of the physiology of production. \ref{fig:PitchEnergyContour} shows temporal trajectories of extracted pitch and energy across a region partly comprised of \textit{taan}, where we clearly observe the pitch modulation and rapid energy fluctuations. There is no apparent correlation between instantaneous values of pitch and vocal energy.  In order to capture the energy fluctuation cue, we use the measured zero-crossing rate from the mean-removed energy contour over 1 s window duration at 500 ms hop.

Next, a local averaging of the features is carried out over 5 s windows to obtain smoothened feature trajectories sampled at 1 s frame rate.  The feature values are normalized to zero-mean and unit variance across the concert. We thus obtain a 3-dimensional normalized feature vector at 1 s frame rate in the vocal segments of the audio which can be used to classify frames into \textit{taan} and non-\textit{taan} categories.

\subsection{Classification and grouping using posteriors}\label{ClassfcntGrpng}
A frame-wise classification into \textit{taan} and non-\textit{taan} styles is carried out for all frames in the vocal segments by a trained MLP network. We use a feed-forward architecture with the sigmoid activation function for the hidden layer comprising 300 neurons. Training uses cross entropy error minimization via the error back-propagation algorithm. Upon classification, the recall and precision of \textit{taan} frame detections with respect to the ground-truth can serve to measure the discriminative power of the features. In our use case however we seek to label continuous regions of the audio rendered in \textit{taan} style much as a human annotator would. This requires the grouping of frames based on homogeneity with respect to the \textit{taan} characteristics. Novelty detection based on a self-distance matrix is an effective way to find segment boundaries \cite{Paulus}. We use a recently proposed approach to computing the SDM from the posterior probabilities derived from the features rather than the features themselves \cite{PV}. The use of Euclidean distance between vectors comprised of posteriors probabilities is found to provide for an SDM with enhanced homogeniety due to the reduced sensitivity to irrelevant local variations. The posteriors are the class-conditional probabilities obtained from the MLP classifier for each test input frame.

Points of high contrast in the SDM are detected by convolution along the diagonal with a checker-board kernel whose dimensions depend upon the desired time scale of segmentation. Considering that the minimum \textit{taan} episode duration, this is chosen to be 5 s in the interest of obtaining reliable boundaries with minimal missed detections. The resulting novelty function is searched for peaks, representing segment boundaries, using `local peak local neighborhood' \cite{turnbulsupervised}. Whether a region between two detected boundaries corresponds to a \textit{taan} is determined by examining the majority of the frame-level classification in that region. Finally, the highest level of grouping is obtained by examining the region of audio separating every two detected \textit{taan} segments. A simple heuristic is set up to mimic the musician’s annotation where \textit{taan} episodes separated by non-\textit{taan} vocal activity of within 20 s are merged into a single section. The merging is also applied if the separation corresponds to a purely instrumental region of duration within 50 s.

\section{Classification with CNN}\label{sec:ClassfcntCNN}
Convolutional Neural Networks are a special case of feed forward neural networks where connections between neurons are restricted to local regions and connection weights are shared. This greatly reduces the model complexity compared to fully connected networks, allowing them to deal with high dimensional inputs such as images or spectrogram excerpts. A CNN consists of convolutional layers, pooling layers and fully connected layers. A convolutional layer computes a convolution of the previous layer outputs with fixed size filter kernels of learnable weights, followed by a non-linear activation function. A convolutional layer consists of multiple such filter kernels producing an output map for each kernel. Convolutional layers are optionally followed by pooling layers which spatially downsample the outputs of the previous layer. The final convolutional or pooling layer of the CNN is typically followed by one or more fully connected layers which reshape the output maps into feature vectors which are finally fed to the output layer.

\subsection{CNN Inputs}
We use excerpts of the spectrograms of our audio files as the input to the CNN. For each of our audio files, sampled at 8 kHz, we compute the log magnitude spectra using a 1024 point DFT on 40 ms Hamming windowed data segments at 20 ms intervals. We believe that the \textit{taan} section can be sufficiently characterized by the temporal variations of the first 2 to 3 vocal harmonics that lie within the frequency range of 0-1.5 kHz. Thus, in order to keep input feature dimension sizes reasonable, we retain only the first 94 frequency bins of the spectrogram corresponding to the frequency range of 0-1469 Hz. We then divide the spectrogram into temporal chunks of 1 s corresponding to our frame size (similar to that used in ground-truth labeling as well as in the hand-crafted features computation). Thus the inputs to the CNN are 94x50 dimensional matrices. By matching the spectrogram resolution and dimensions to our task, we eliminate the need for multiple channel inputs as has been the case in a previous audio task \cite{ImprovedOnsetCNN}. To bring the input values within a suitable range, we normalize each frequency band to zero mean and unit variance using the mean and standard deviation values estimated using the training. \cite{ImprovedOnsetCNN}.

\subsection{CNN Architecture}
\begin{figure}[t]
 \centerline{\framebox{
 \includegraphics[width=\columnwidth,height=4cm]{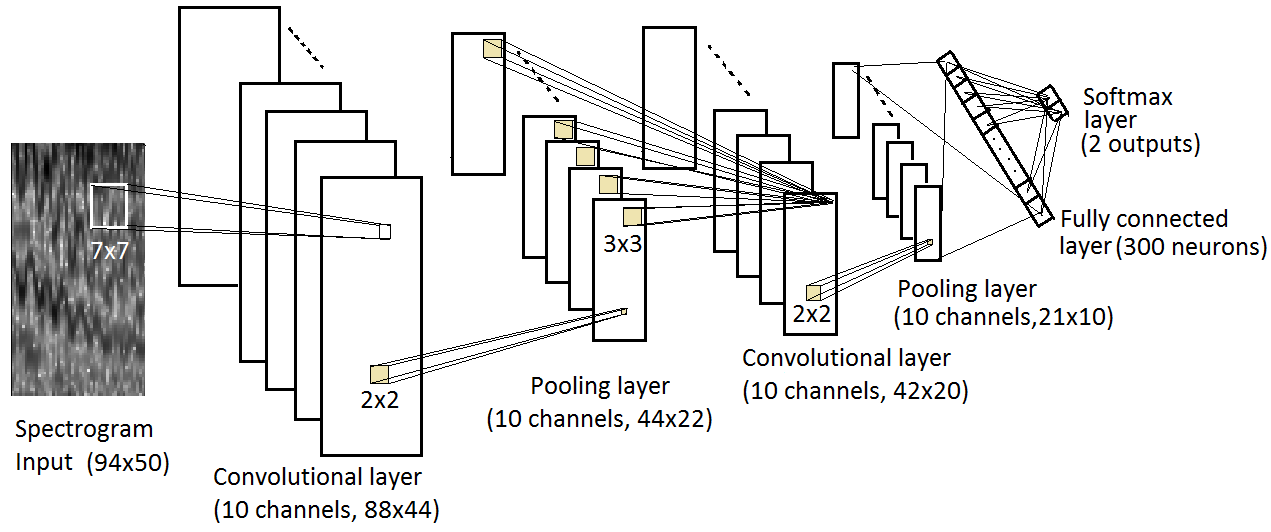}}}
 \caption{The CNN architecture employed}
 \label{fig:CNNArchtctr}
\end{figure}

The Convolutional Neural Network used in this work has an architecture similar to that described in \cite{schluterOnsetCNN}; the main difference being that the input spectrogram excerpts in our case use a single time resolution as opposed to having multiple input channels with different time resolutions in \cite{schluterOnsetCNN}. Our network architecture is summarized in \ref{fig:CNNArchtctr}. The CNN has five layers in total, two convolutional layers, two pooling layers and a fully connected layer. The first layer of the network is a convolutional layer consisting of 10 7x7 filter kernels producing 10 output maps of size 88x44 each. This is followed by an average pooling layer which retains the average value of non-overlapping 2x2 cells. This is followed by another convolutional layer of 10 3x3 filters and another 2x2 average pooling layer to give 10 output maps of size 21x10. These outputs are then reshaped to a 2100 dimensional feature vector and fully connected to a layer of 300 sigmoidal units. The outputs of these 300 units are finally given to a softmax output layer consisting of 2 units corresponding to the two classes being considered.

\subsection{Training the CNN}
The CNN training is carried out in two stages. The CNN is first trained without the fully connected layer, with the outputs of the second pooling layer directly connected to the softmax layer. The outputs of the trained CNN at the second pooling layer are then concatenated into a feature vector for each frame of the training data. These feature vectors are then treated as the training data for a Multi Layer Perceptron network with a single hidden layer of 300 units and a softmax output layer of two units. Finally, the trained CNN and MLP together form the CNN with the fully connected layer. The CNN and MLP are trained using the Error Backpropagation algorithm for minimizing the cross-entropy error between the softmax outputs and the labels for each 1 sec input frame of the training data. Training is carried out for a fixed 900 epochs over the train set of 35 concerts as described in section \ref{Database}. An initial learning rate of 0.1 is halved after every 150 epochs.

\section{Experiments and Evaluation}
Our ideal system would detect and segment \textit{taan} sections similar to a musician’s labeling. This high level task is attempted by the sequence of frame-level automatic classification and higher level grouping as described in section \ref{sec:FeatExtrClasGrpng}. In this section, we present experimental results on the performance of each of the components. Frame-level classification is measured by the detection of \textit{taan} in terms of recall and precision. Artist-dependent and artist-independent training are compared for the hand-crafted features based classifier. The same evaluations are carried out with the CNN classifier where the “features” are purely learned during training.

The frame-level classification needs frame-level (i.e. 1 s resolution) annotation of \textit{taan} presence or absence. This is required both for the training of the classifiers as well as for reliable testing. The musician labels are not useful as such for this end due to the presence of non-\textit{taan} interruptions of significant duration within the musician labeled \textit{taan} sections as seen in \ref{fig:PraatSpecImg}.  Thus, for the development of the frame-level classifier, we need a more fine-grained marking of \textit{taan} segments. Since this is a demanding task to carry out manually, we use a bootstrapped iterative approach where, for each concert audio, a 2-mixture GMM on the melodic style feature vector is fitted to a small amount of hand labeled data and updated with classified frames across the audio track in each iteration until convergence is achieved \cite{nguyen}. Casual inspection showed that the frame-level labels so obtained were indeed accurate and these were then used to train and evaluate the frame level classifiers.

\begin{figure}[t]
 \centerline{\framebox{
 \includegraphics[width=\columnwidth]{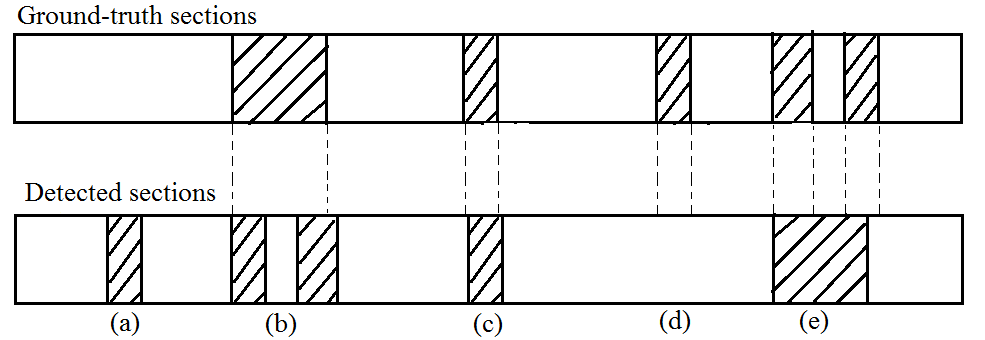}}}
 \caption{Various scenarios that occur after grouping viz. (a) false alarm, (b) over-segmentation, (c) exact detection, (d) missed detection, (e) under-segmentation}
 \label{fig:EvalMetric}
\end{figure}

The system is also evaluated after grouping, this time in terms of the match between the detected segments and the subjectively labeled \textit{taan} segments for each concert. Measures of performance include the number of correctly retrieved \textit{taan} segments and number of false alarms. A section is said to be correctly retrieved if there is an overlap of at least 50\% of its duration with a detected segment.  Also of interest is the extent of over- or under-segmentation of the correctly detected \textit{taan} sections. \ref{fig:EvalMetric} illustrates the different possibilities of mismatch that are observed between subjective labels and automatically labeled sections. When subjectively labeled section is correctly detected, it is observed that the onset and offset boundaries are always within 5 s of the corresponding ground-truth boundaries indicating the reliability of the posteriors based segmentation.

\section{Results and Discussion}\label{RnD}
As mentioned in section \ref{Database}, our experimental evaluation of the two different frame-level classifier systems is based on (i) a single-artist concert dataset of 22 concerts trained and tested in leave-one-concert-out cross-validation mode, and (ii) testing on the same 22 concerts but with training on a large dataset where the given artist is not represented. \ref{ref:ROC22J3522V} (a) and (b) show the ROCs corresponding to each of these train-test scenarios. We observe that the performance of the hand-crafted features is superior to that obtained by the CNN in each case. We present some insights related to this in the next section. 

\begin{figure}[t]
 \centerline{\framebox{
 \includegraphics[width=\columnwidth]{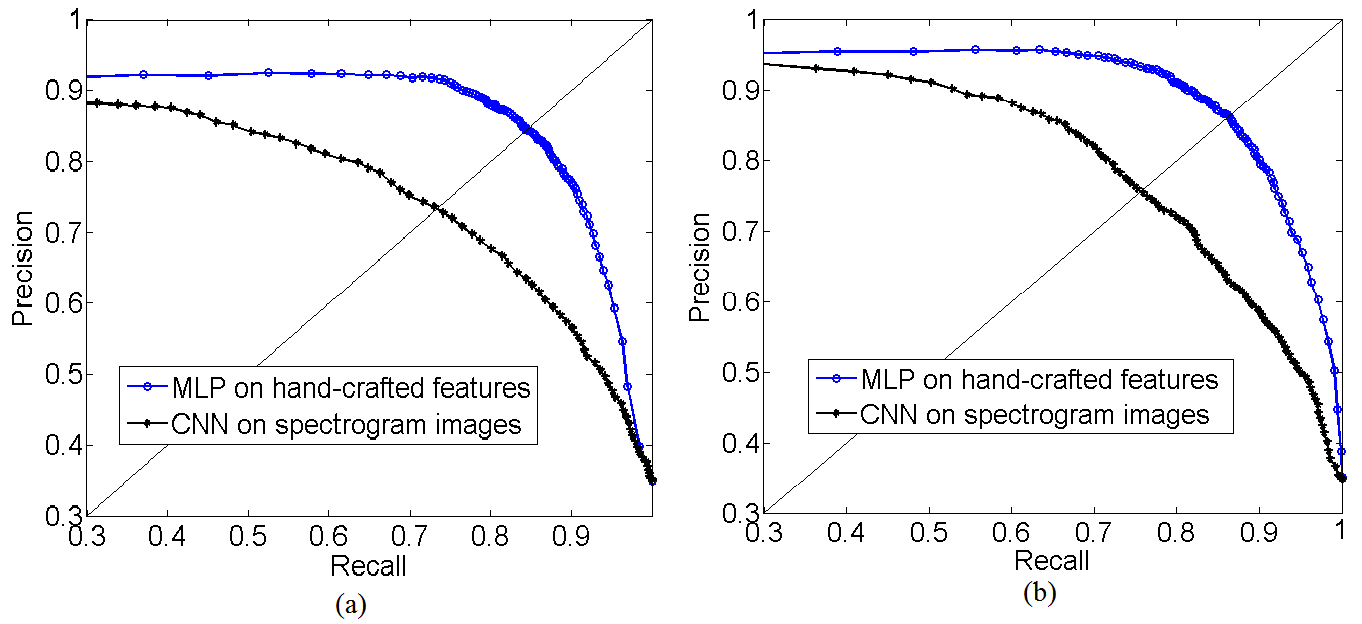}}}
 \caption{(a)ROC for CNN and MLP on hand-crafted features for leave-one-song-out case of 22 concerts. (b)ROC for CNN and MLP on hand-crafted features for 35 train and 22 test concert scenario.}
 \label{fig:ROC22J3522V}
\end{figure}

By noting the equal error rates (precision = recall) for each classifier across the training sets, we see that the performance improves when the training dataset size is increased from 22 concerts to 35 concerts (which in reality is a 3-fold increase in the number of labeled frames in the training data due to the longer concert durations). Thus it appears that any gains from intra-artist training are over-shadowed by the benefits from the larger training data size. This could also be related to the fact that \textit{taan} singing style characteristics are relatively artist independent.

The second stage of frame grouping and segmentation is implemented with the frame-level posteriors obtained by the MLP classifier using the hand-crafted features operating at its optimal operating point (f score = 0.86). Using the method presented in section \ref{sec:FeatExtrClasGrpng}, we obtain the results provided in \ref{tab:EvaluationST2}. We note that of the 115 subjectively labeled \textit{taan} sections across the 22 concerts, there are only 9 missed detections. We have 2 false detections. We thus have a system that does indeed accurately flag the occurrence of \textit{taan} sections across concerts. Finally of the 106 correct detections, the majority are correctly segmented. Over- and under-segmentations account for a third of the detections. These can possibly be corrected by modifying the heuristics of the highest level of grouping (bridging over gaps) discussed in section \ref{sec:FeatExtrClasGrpng}. Deriving empirical rules regarding high-level segmentation by musicians would ideally require a study over a larger database with more human annotators per concert.

\begin{table}[t]
 \begin{center}
\begin{tabular}{|c|l|l|}
\hline
{\begin{tabular}[c]{@{}c@{}}True Detection\\ (106)\end{tabular}} & Under-Segmentation & 32 \\ \cline{2-3} 
                                                                              & Over-Segmentation  & 3  \\ \cline{2-3} 
                                                                              & Exact Detection    & 71 \\ \hline
\multicolumn{2}{|l|}{Missed}                                                                       & 9  \\ \hline
\multicolumn{2}{|l|}{False Alarm}                                                                  & 2  \\ \hline
\end{tabular}
\end{center}
 \caption{Segmentation performance after grouping}
 \label{tab:EvaluationST2}
\end{table}

\section{Some Insights}
While we noted in the previous section that the hand-crafted features perform better than the CNN learned features, it is interesting to look deeper at the distribution of frame-level errors shown in \ref{tab:McNemar}. We note that while the CNN features misclassify more frames in total, there are also a sizeable number of frames that are misclassified by the hand-crafted features but correctly classified by the CNN. This indicates the presence of complementary information and that a combination of classifiers is very likely to yield a performance superior to any one of the systems.

\begin{table}[t]
 \begin{center}
\begin{tabular}{llll}
                                                                                                       &                                & \multicolumn{2}{c}{CNN}                                       \\ \cline{3-4} 
                                                                                                       & \multicolumn{1}{l|}{}          & \multicolumn{1}{l|}{Correct} & \multicolumn{1}{l|}{Incorrect} \\ \cline{2-4} 
\multicolumn{1}{l|}{{\begin{tabular}[c]{@{}l@{}}Hand-crafted \\ features\end{tabular}}} & \multicolumn{1}{l|}{Correct}   & \multicolumn{1}{l|}{4998}    & \multicolumn{1}{l|}{762}       \\ \cline{2-4} 
\multicolumn{1}{l|}{}                                                                                  & \multicolumn{1}{l|}{Incorrect} & \multicolumn{1}{l|}{272}     & \multicolumn{1}{l|}{296}       \\ \cline{2-4} 
\end{tabular}
\end{center}
 \caption{Distribution of classification errors}
 \label{tab:McNemar}
\end{table}

\begin{figure}[t]
 \centerline{\framebox{
 \includegraphics[width=\columnwidth]{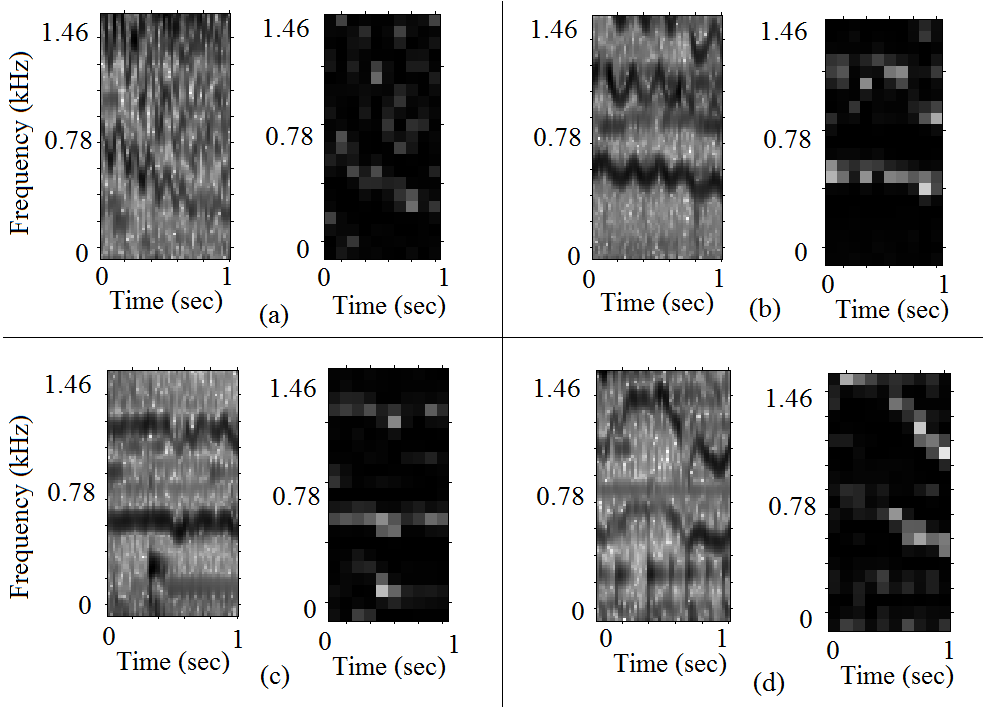}}}
 \caption{Input spectrograms and $9^{th}$ channel output maps for the $2^{nd}$ pooling layer. (a) Correctly classified as \textit{taan} by both CNN and hand-crafted features (b) Incorrectly classified as non-\textit{taan} by CNN and correctly classified as \textit{taan} by hand-crafted features (c) Correctly classified as non-\textit{taan} by both CNN and hand-crafted features (d) Incorrectly classified as \textit{taan} by CNN but correctly classified as non-\textit{taan} by hand-crafted features}
 \label{fig:CNNInputOutput}
\end{figure}

The hand-crafted features were designed to capture the temporal modulation of the pitch and energy trajectories after suitable normalization steps. This information is, of course, implicitly encoded in the spectrogram via the first several strong harmonics of the vocal source.  Our choice of spectrogram parameters at the input of the CNN makes the same information, at least in spatial image form, available to the convolutional layers. We select a few examples to obtain an understanding of the encoding of \textit{taan} and non-\textit{taan} distinctions by the CNN features. In order to study the learned features, we note that the outputs of the second pooling layer finally get concatenated to form the feature vector for classification. Since the second pooling layer is the last layer where the outputs show spatial correspondences with the input spectrogram image, observing the outputs of the second pooling layer could give us insight into what the CNN encodes in each image. \ref{fig:CNNInputOutput} shows the input spectrogram patches for four different frame categories (based on classification achieved by each of the two systems) and the corresponding outputs at the $9^{th}$ channel of the second pooling layer. The $9^{th}$ channel was one of the channels having larger connection weights to the fully connected layer compared to other channels, implying that its outputs were more significant than those of the other channels for the classification.

From \ref{fig:CNNInputOutput} we observe that the outputs of the second pooling layer are rather sparse with respect to the input spectrograms, indicating that the high level features learned by the CNN may be discarding the less relevant parts of the input spectrogram.  Here the retained structure seems to correspond to higher energy portions of the spectrogram such as the vocal harmonics and occasionally other instrumental harmonics and percussion strokes. \ref{fig:CNNInputOutput}(b) and (c) show frames that were classified as non-\textit{taan} by the CNN. The frame in \ref{fig:CNNInputOutput}(c) actually corresponds to a non-\textit{taan} frame characterised by its non oscillating almost constant vocal harmonics, which did get captured as horizontal lines. The frame in \ref{fig:CNNInputOutput}(b), however was actually a \textit{taan} frame as seen by the oscillating vocal harmonics. However, the oscillations in the  first harmonic at about 600 Hz were not prominent enough and got captured as a virtually horizontal structure leading to the misclassification. \ref{fig:CNNInputOutput} (a) and (d) show frames that were classified as \textit{taan} by the CNN. \ref{fig:CNNInputOutput}(a) was indeed a \textit{taan} frame. The rapid oscillations in its vocal harmonics appear as a scattered pattern in the output map. \ref{fig:CNNInputOutput}(d) represents a common CNN misclassification. This non-\textit{taan} frame has time-varying harmonics but the time-variation is not a regular pitch modulation characteristic of \textit{taan}. The output map shows a breakdown of the harmonic structure indicating that the precise nature of the time variation is not learned by the CNN features. Rather, the CNN appears to characterize non-\textit{taan} frames, which are marked by the presence of stable or at most slow varying vocal harmonics, with near horizontal lines in the output maps, and all inputs that do not match these stable characteristics as \textit{taan}.

Finally, we also examined cases where the CNN features correctly classified \textit{taan} frames that were missed by the hand-crafted features. These frames had spectrogram images that clearly showed the oscillating harmonics. However it turned out that pitch tracking errors in these frame led to the loss of this information capture in the hand-crafted features. This raises the important point that the learning from raw audio spectra via the CNN could decrease vulnerability to errors in fixed high-level feature extraction modules such as predominant pitch detection.

\section{Conclusion}
We proposed a system for the segmentation and labeling of a prominent named structural component of the Hindustani vocal concert. The \textit{taan} section is characterized by a melodic style marked by rapid pitch and energy modulation of the singing voice. High-level features to capture this specific modulation from the pitch tracks extracted from the polyphonic audio, combined with novelty based grouping of frame posteriors, provided high accuracy \textit{taan} segmentation on our test dataset of concerts. We also investigated the possibility of automatically learning distinctive features, using a CNN for this task, from raw magnitude spectra computed from the polyphonic audio signal. Notwithstanding that we approached this particular comparison with a healthy dose of skepticism, it was observed that the CNN did indeed perform the frame-level classification far better than chance. An inspection of the outputs of the second pooling layer reflected a systematic difference in \textit{taan} and non-\textit{taan} frames. Although non-\textit{taan} frames where the harmonics varied over time were misclassified as \textit{taan} frames, it is entirely possible that training on a larger dataset with more such instances as well as using a network with more layers could help improve performance. Finally, the complementary errors of the two classifier systems can lead to fruitful combinations for further improvements in performance. The more general conclusion is that learned features can indeed add value to hand-crafted features in audio retrieval tasks.

\section{Acknowledgement}
This work received partial funding from the European Research Council under the European Union’s Seventh Framework Programme (FP7/2007-2013)/ERC grant agreement 267583 (CompMusic).

{\small
\bibliographystyle{ieee}
\bibliography{egbib}

\begin{thebibliography}{10}

\bibitem{Praat}
P.~Boersma and D.~Weenink.
\newblock Praat, a system for doing phonetics by computer.
\newblock {\em Glot International}, 5(9/10):341--345, 2001.

\bibitem{ChitraSpringer}
C.~Gupta and P.~Rao.
\newblock {\em Objective Assessment of Ornamentation in Indian Classical
  Singing}, volume 7172 of {\em Lecture Notes in Computer Science}.
\newblock Springer Berlin Heidelberg, 2012.

\bibitem{ChordRecogBello}
E.~Humphrey and J.~Bello.
\newblock Rethinking automatic chord recognition with convolutional neural
  networks.
\newblock In {\em Proceedings of 11th International Conference on Machine
  Learning and Applications}, volume~2, pages 357--362, Dec 2012.

\bibitem{AndrewCNN}
H.~Lee, P.~Pham, Y.~Largman, and A.~Ng.
\newblock Unsupervised feature learning for audio classification using
  convolutional deep belief networks.
\newblock In {\em Proceedings of Advances in neural information processing
  systems}, pages 1096--1104, 2009.

\bibitem{nguyen}
T.~Nguyen, H.~Sun, SK. Zhao, SZK. Khine, HD. Tran, TLN. Ma, B.~Ma, ES. Chng,
  and H.~Li.
\newblock The iir-ntu speaker diarization systems for rt 2009.
\newblock In {\em RT’09, NIST Rich Transcription Workshop}, volume~14, pages
  17--40, 2009.

\bibitem{Paulus}
J.~Paulus, M.~Muller, and A.~Klapuri.
\newblock State of the art report: Audio based music structure analysis.
\newblock In {\em Proceedings of the International Symposium on Music
  Information Retrieval}, pages 625--636, 2010.

\bibitem{rao2014overview}
S.~Rao and P.~Rao.
\newblock An overview of hindustani music in the context of computational
  musicology.
\newblock {\em Journal of New Music Research}, 43(1):24--33, 2014.

\bibitem{VRaoContextAware}
V.~Rao, C.~Gupta, and P.~Rao.
\newblock Context-aware features for singing voice detection in polyphonic
  music.
\newblock In {\em Proceedings of of Adaptive Multimedia Retrieval}, pages
  43--57, 2013.

\bibitem{VRaoMelodyExtract}
V.~Rao and P.~Rao.
\newblock Vocal melody extraction in the presence of pitched accompaniment in
  polyphonic music.
\newblock {\em Audio, Speech, and Language Processing, IEEE Transactions on},
  18(8):2145--2154, Nov 2010.

\bibitem{schluterOnsetCNN}
J.~Schl{\"u}ter and S.~B{\"o}ck.
\newblock Musical onset detection with convolutional neural networks.
\newblock In {\em 6th International Workshop on Machine Learning and Music,
  Prague, Czech Republic}, 2013.

\bibitem{ImprovedOnsetCNN}
J.~Schl{\"u}ter and S.~B{\"o}ck.
\newblock Improved musical onset detection with convolutional neural networks.
\newblock In {\em Proceedings of IEEE International Conference on Acoustics,
  Speech and Signal Processing}, pages 6979--6983, 2014.

\bibitem{XSerra}
X.~Serra.
\newblock Exploiting domain knowledge in music information research.
\newblock In {\em Proceedings of Stockholm Music Acoustics Conference and Sound
  and Music Computing Conference}, pages 3--6, 2013.

\bibitem{turnbulsupervised}
D.~Turnbull, G.~Lanckriet, E.~Pampalk, and M.~Goto.
\newblock A supervised approach for detecting boundaries in music using
  difference features and boosting.
\newblock In {\em Proceedings of the International Symposium on Music
  Information Retrieval}, 2007.

\bibitem{Ullrich}
K.~Ullrich, J.~Schlüter, and T.~Grill.
\newblock Boundary detection in music structure analysis using convolutional
  neural networks.
\newblock In {\em Proceedings of the International Symposium on Music
  Information Retrieval}, 2014.

\bibitem{PV}
P.~Verma, T.~P. Vinutha, P.~Pandit, and P.~Rao.
\newblock Structural segmentation of hindustani concert audio with posterior
  features.
\newblock In {\em Proceedings of IEEE International Conference on Acoustics,
  Speech and Signal Processing}, 2015.

\end{thebibliography}
}

\end{document}